\documentclass{article}
\usepackage{graphicx,amssymb}
\pdfoutput=1
\usepackage[colorlinks=true,urlcolor=blue,
citecolor=blue,linkcolor=blue,pdfa=true]{hyperref}
\begin{document}
\title{\bfseries
	Early Results from TUS, the First Orbital Detector of Extreme
	Energy Cosmic Rays\thanks{Written version of a talk given at
	\href{https://indico.cern.ch/event/504078/}{UHECR2016}, Kyoto, 11--14 October, 2016}}

\author{Mikhail Zotov for the Lomonosov--UHECR/TLE Collaboration\\
	\itshape
	M.V.~Lomonosov Moscow State University,\\
	\itshape
	D.V.~Skobeltsyn Institute of Nuclear Physics,\\
	\itshape
	Moscow, Russia
	}
\date{}
\maketitle

\begin{abstract}
	TUS is the world's first orbital detector of
	extreme energy cosmic rays (EECRs), which operates as a part of the
	scientific payload of the Lomonosov satellite since May 19, 2016.
	TUS employs the nocturnal atmosphere of the Earth to
	register ultraviolet (UV) fluorescence and Cherenkov radiation from
	extensive air showers generated by EECRs as well as
	UV radiation from lightning strikes and transient luminous events,
	micro-meteors and space debris.
	The first months of its operation in orbit have demonstrated an
	unexpectedly rich variety of UV radiation in the atmosphere.
	We briefly review the design of TUS and present a few examples of
	events recorded in a mode dedicated to registering EECRs.
\end{abstract}

\section{Introduction}

Extreme energy cosmic rays (EECRs) with energies $\gtrsim50$~EeV were
discovered more than 50~years ago~\cite{Linsley-1e20-1963}, but their
origin and nature still remain unclear, see~\cite{Dawson_etal-2017} for
the latest review.  To a great extent, the problem relates to the very
low flux of EECRs.  It is sufficient to say that the largest existing
experiment, the Pierre Auger Observatory (Auger for short), registered
only 146 cosmic rays with energies above 53~EeV in nearly eight years of
operation~\cite{2015ApJ...804...15A}.  A smaller Telescope Array (TA)
experiment obtained 83 events above 57~EeV in seven
years~\cite{Tinyakov-ICRC-2015}.  Another difficulty arises from the
fact that neither of the two installations has a full coverage of the
celestial sphere.

Both experiments employ a so-called hybrid technique of registering
cosmic rays. It includes an array of surface detectors placed on a large
area (3000~km$^2$ in case of Auger and approximately 700~km$^2$ for
the TA) and a set of fluorescence telescopes aimed to register ultraviolet
(UV) fluorescent and Cherenkov radiation emitted by ionized molecules of
nitrogen excited by charged particles of extensive air shower (EAS)
cascades generated in the atmosphere by cosmic rays.  Another approach
aimed to drastically increase the exposure of an experiment was put
forward by Benson and Linsley, who suggested to register both kinds of
UV radiation emitted during the development of an EAS with an orbital
telescope~\cite{1980BAAS...12Q.818B,Benson-Linsley-1981}.
They estimated that a huge circular field of view about 100~km in
diameter, three times larger than the coverage of Auger, was possible
with a satellite flying on a circular equatorial orbit at height
500--600~km and equipped with a mirror 36~m in diameter with~$10'$
resolution and $\sim5000$ photomultiplier tubes located
at the focal surface of the mirror, at a duty cycle of the order of
20--30\%, and the energy threshold below 10~EeV.

While the idea might look simple, one has to overcome a whole number of
scientific and technical difficulties to implement it.  Despite of a
number of efforts, none of the projects was implemented until April~28,
2016, when TUS (Tracking Ultraviolet Set-up), the first orbital
detector of EECRs, was launched into orbit from the newly built
Vostochny Cosmodrome (Russia) as a part of the scientific payload of the
Lomonosov satellite (international designation MVL~300, or 2016-026A).
In what follows, I outline the design of TUS and briefly
present some of the early results of the on-going analysis of data obtained
during the first months of TUS' operation in orbit.

\section{Design of TUS}

The idea by Benson and Linsley attracted attention of physicists in 
Russia in late 1990s.
TUS was first announced in 2001 as a pathfinder for a more advanced
KLYPVE mission~\cite{2001ICRC....2..831A}.
Remarkably, John~Linsley participated in the project at its early stage.
The development was led by the D.V.~Skobeltsyn Institute of
Nuclear Physics at M.V.~Lomonosov Moscow State University in
collaboration with a number of universities and research organizations
in Russia, Korea, and
Mexico~\cite{TUS-ecrs2012,TUS-expastron}.
TUS inherited the optical scheme of the original design by Benson and
Linsley but with modest technical parameters. The instrument consists
of a parabolic mirror-concentrator of the Fresnel type and a
square-shaped $16\times16$-channel photodetector aligned to the focal
plane of the mirror.  The mirror has an area of about 2~m$^2$ and a
1.5~m focal distance.  Pixels of the photodetector are Hamamatsu R1463
photomultiplier tubes (PMTs) with a 13~mm diameter multialkali cathode.
The pixels are grouped in 16 identical photodetector modules.
Light guides with square entrance apertures and circular outputs are
employed to uniformly fill the focal surface.  All PMTs have black
blends extending 1~cm above their light guides as a kind of protection
against side illumination.  A 2.5~mm thick UV filter is placed in front
of each PMT cathode to limit the measured wavelength to the 300--400~nm
range.  The field of view (FOV) of TUS equals $\pm4.5^\circ$, which
results in an area of approximately 80~km$\times$80~km at sea level so
that a single channel observes a 5~km$\times$5~km
square~\cite{2014NIMPA.763..604G}.
(The Lomonosov satellite has a sun-synchronous orbit with an inclination
of $97^\circ\!\!.3$, a period of $\approx94$~min, and a height of about
470--500~km.)

TUS can operate in four modes with different time sampling windows
(``time frames'').
The main mode is aimed at registering EASs and has a time sampling
window of 0.8~$\mu$s.  Time frames of 25.6~$\mu$s and 0.4~ms
are utilized for studying transient luminous events (TLEs) of different
kinds, and a window of 6.6~ms is available for detecting micro-meteors
and possibly space debris.
A data record of any TUS event includes 256 discrete waveforms, one for
each channel, and every waveform contains analog-to-digital converter
(ADC) counts for 256 time frames.  Due to certain limitations of the
Lomonosov hardware, all parameters of the trigger system are adjusted so
that the trigger rate does not exceed approximately one event
per minute.  All four modes of operation were tested since the launch.

Recall that the situation with the energy spectrum of cosmic rays
above $\sim50$~EeV was unclear at the time the TUS project was put forward.
In particular, it was suggested by the Haverah
Park~\cite{1991JPhG...17..733L} and AGASA~\cite{AGASA-spectrum}
experiments that there was no cut-off of the spectrum at energies
predicted by Greisen, Zatsepin and Kuz'min~\cite{Greisen-1966,ZK-1966}.
It was estimated based on the AGASA data that TUS would be able to
register about 20 events with energies above 200~EeV and $\simeq900$
events above 20~EeV in three years of
operation~\cite{2001ICRC....2..831A}.  It became clear that these
expectations were too optimistic as soon as HiRes and Auger found that
the flux of cosmic rays was strongly suppressed at energies
$\gtrsim50$~EeV~\cite{HiRes-GZK-2007,Auger-GZK}. Currently, the main
scientific objective of TUS in relation to cosmic rays is to test the
technique of observing EECRs from space and to obtain data on the UV
background of nocturnal atmosphere to be used in the development of the
future missions such as KLYPVE (\mbox{K-EUSO})~\cite{klypve-2015} and
JEM-EUSO~\cite{JEM-EUSO-tool}.

Still, there are chances that TUS registers a number of EECRs.
Various simulations were performed for TUS before the launch, and the
energy threshold for effective registering of EECRs was determined to be
approximately 70~EeV, so that the instrument should be able to register
several events above the threshold in 5 years of continuous
operation assuming the energy spectrum obtained with
Auger for highly inclined EAS~\cite{TUS-sim-ApP-2017}.
In what follows, we only discuss results obtained in the EAS mode.
A few examples of data recorded in the TLE and meteor modes can be found
in~\cite{TUS-TEPA-2016,TUS-ECRS-2016}.

\section{Selected Results}

\subsection{Instant Track-like Flashes}

\begin{figure}[!ht]
	\begin{center}
		\includegraphics[width=.48\textwidth]{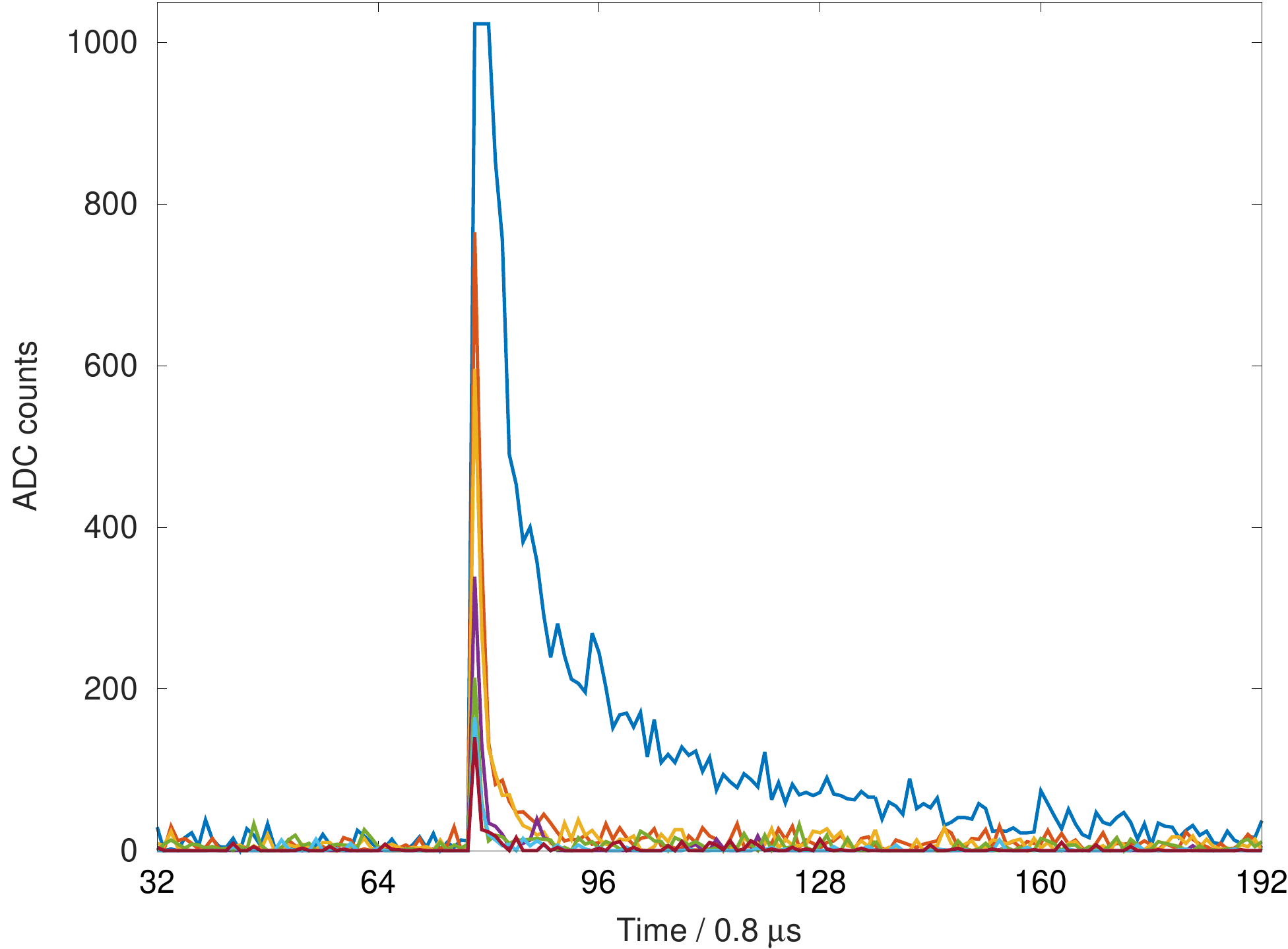}\qquad
		\includegraphics[width=.43\textwidth]{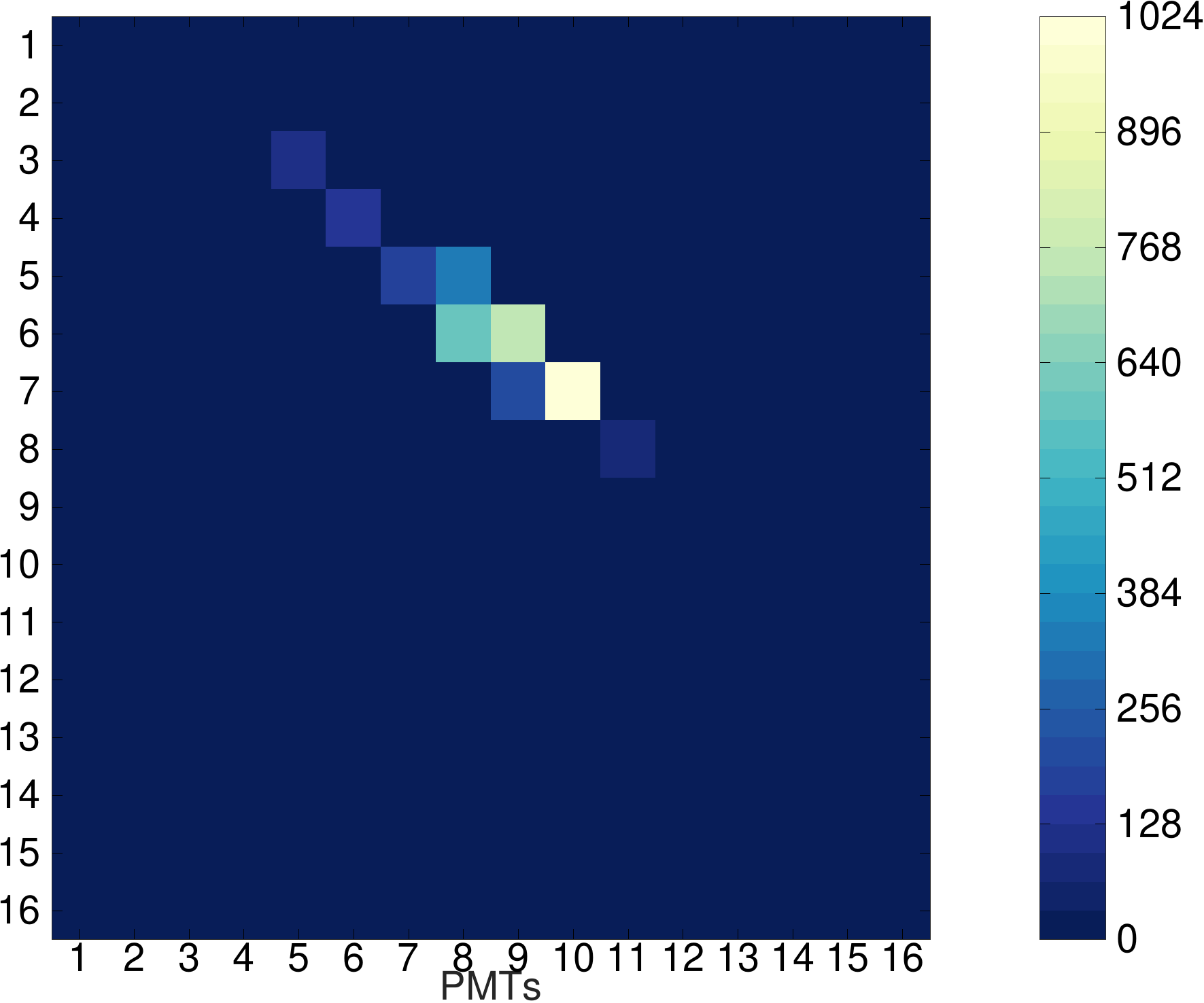}
	\end{center}
	\caption{Example of an instant track-like flash.
		Left: waveforms of a few PMTs with the highest ADC counts.
		Colours denote waveforms of different PMTs.
		Right: snapshot of the focal surface at the moment of the
		maximum signal. Here and below, colours denote ADC counts.
	}
	\label{fig:track}
\end{figure}

The biggest surprise of the first days of the TUS operation in orbit
were multiple strong flashes growing in a single time frame (0.8~$\mu$s,
i.e., instantly in the TUS time scale) simultaneously in a number of
adjacent PMTs usually producing linear tracks in the focal surface.  Due
to these features, they are tentatively called ``instant track-like
flashes.'' An example of such an event is shown in Fig.~\ref{fig:track}.
Notice one of the pixels is saturated reaching the maximum possible ADC
count of 1023, which approximately corresponds to 2000 photons at the
PMT pupil in conditions of a low background flux.  The situation is
typical for this kind of flashes, see other examples
in~\cite{TUS-TEPA-2016,TUS-ECRS-2016,tracks-izvran}.  The length and shape
of tracks produced in the focal surface vary from one event to
another but we have not found any correlation of them with time of observation
or position of Lomonosov yet, though waveforms demonstrate some specific
features on certain days.  Total intensity of the flashes also varies
with the number of saturated pixels exceeding eight in some cases.  The
flashes constitute 12\% of nearly 25 thousand events recorded in the EAS
mode from May~19 till December~16, 2016.  Their geographic distribution
has two distinct maxima at latitudes $35^\circ$S--$50^\circ$S and
$45^\circ$N--$60^\circ$N and a minimum at $0^\circ$--$15^\circ$N.  This
is drastically different from the distribution of the whole data set,
which is approximately uniform in latitudes $50^\circ$S--$70^\circ$N.

The instant track-like flashes cannot result from extensive air showers
generated by EECRs in the atmosphere because a nearly horizontal EAS
should produce a track composed of adjacent pixels that flash throughout
approximately 16 consecutive time frames, not all at once.  Simulations
performed using the Geant4 software toolkit~\cite{Geant4} have revealed
that protons with energies from 100--200~MeV up to a few GeV that hit
and penetrate the UV filters nearly parallel can produce UV radiation,
and result in tracks similar to those recorded by
TUS~\cite{tracks-izvran}.  On the other hand, weak flashes with low ADC
counts and only one or a few pixels involved are potentially originated
from electrons of the inner radiation belt.
Other possible sources of this kind of events are also being considered.  

According to other simulations, Cherenkov light, either reflected from
Earth or generated within the mirror medium by an upward-going particle
can also potentially generate instant flashes with a spot produced in
the focal plane~\cite{2011BRASP..75..381S,2014AdSpR..53.1515B}.  The
probability of this kind of events is small, but its importance calls
for a dedicated analysis.

\subsection{Flashes Related to Thunderstorms}

\begin{figure}[!ht]
	\begin{center}
		\includegraphics[width=.48\textwidth]{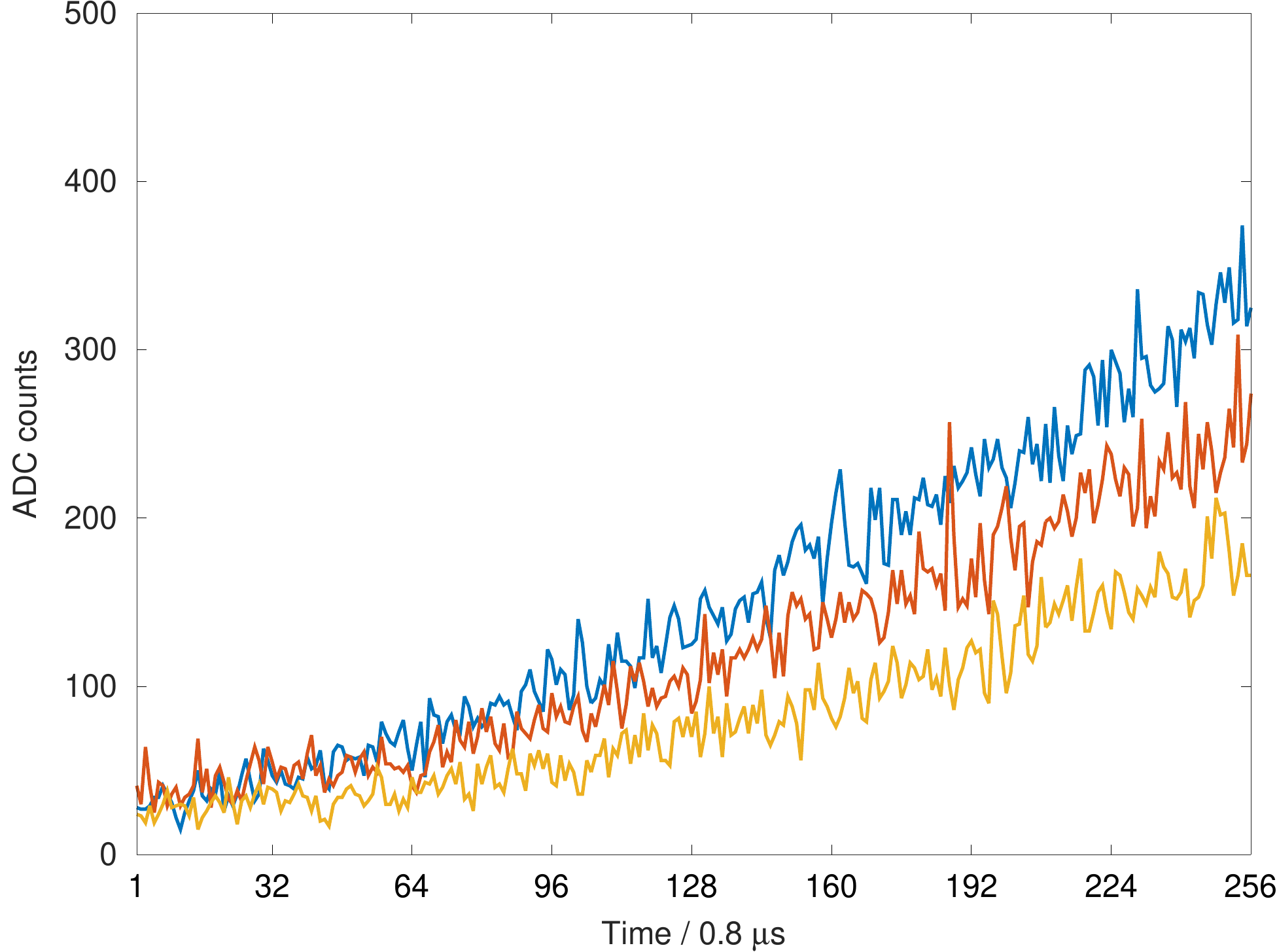}\qquad
		\includegraphics[width=.43\textwidth]{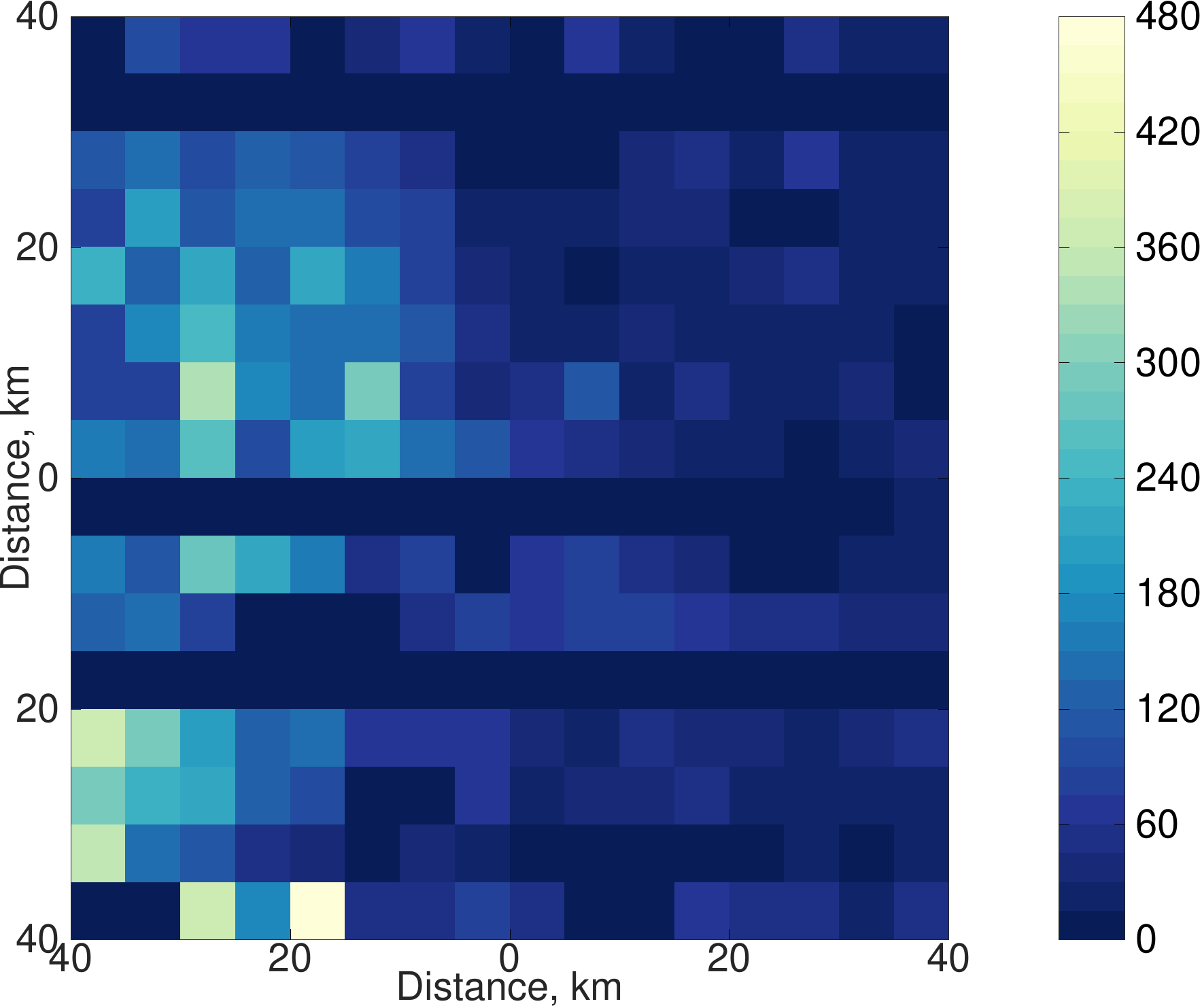}
	\end{center}
	\caption{Example of a slow flash registered on September~12, 2016,
		at 04:20:09 UTC. The centre of the FOV was at $21^\circ\!\!.8$N,
		$74^\circ\!\!.5$W.
		Left: waveforms of three PMTs.
		Right: snapshot of the focal plane at the maximum of the flash.
		Here and below,
		numbers to the left and beneath snapshots indicate
		approximate distances from the centre of the FOV at
		ground level;
		geographic North is at $11^\circ$--$12^\circ$ counterclockwise
		from the top of the focal plane,
		East is respectively to the right.
		Rows with zero ADC counts are due to malfunctioning PMT modules.
		}
	\label{fig:slowflash}
\end{figure}

Another interesting group of events consists of spatially extended
flashes with ADC counts that monotonically increase (up to random
fluctuations of the signal) during $\gtrsim60~\mu$s as shown in the left
panel of Fig.~\ref{fig:slowflash}.  We tentatively call them ``slow
flashes'' to distinguish from the events discussed in the previous
section.  A slow flash typically evolves simultaneously across the focal
surface, producing its nearly uniform illumination.  In most cases, ADC
counts continue growing until the end of a record but sometimes a
maximum is reached and the signal begins vanishing before the record is
complete.  In a few cases, illumination of the focal plane by a slow
flash is strongly non-uniform, as demonstrated in the right panel of
Fig.~\ref{fig:slowflash}.

An analysis of the geographic distribution of slow flashes demonstrated
its correlation with known regions of high lightning flash rates. This
provided a clue to a possible origin of this kind of events.
A sample of slow flashes registered from August~16, 2016, to
September~19, 2016, was compared with data from the World-Wide Lightning Location
Network (WWLLN).
We analysed data at distances up to $\sim2000$~km from the TUS FOV,
which corresponds to the maximum distance from which a ray of light
tangent to the Earth is visible form the Lomonosov orbit.
It was found that for the time window of~$\pm1$~s,
which matches the accuracy of the TUS trigger time stamps, the majority of
``companion'' lightnings were registered at distances 400--1600~km from
positions of the respective flashes recorded by TUS.  It is likely that
the origin of slow flashes uniformly illuminating the focal surface is
diffusive light of distant lightning strikes scattered by the surface of
the TUS mirror.  
In contrast to this situation, a lightning strike was registered by the
WWLLN 0.37~s prior to the TUS trigger in around 85~km South-West from
the centre of the FOV resulting in non-uniform illumination of
the focal plane shown in Fig.~\ref{fig:slowflash}.

\begin{figure}[!ht]
	\begin{center}
		\includegraphics[width=.48\textwidth]{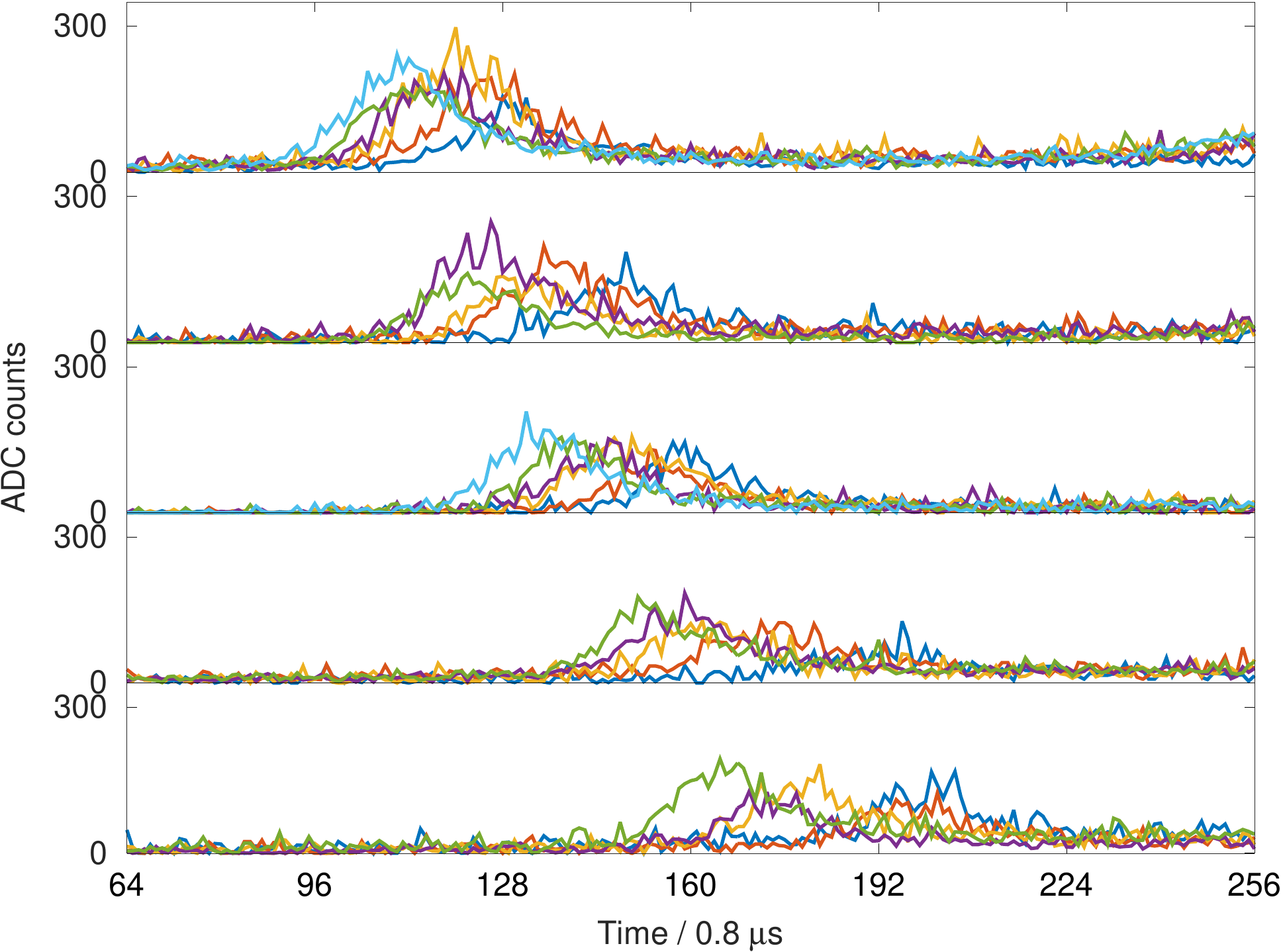}\qquad
		\includegraphics[width=.43\textwidth]{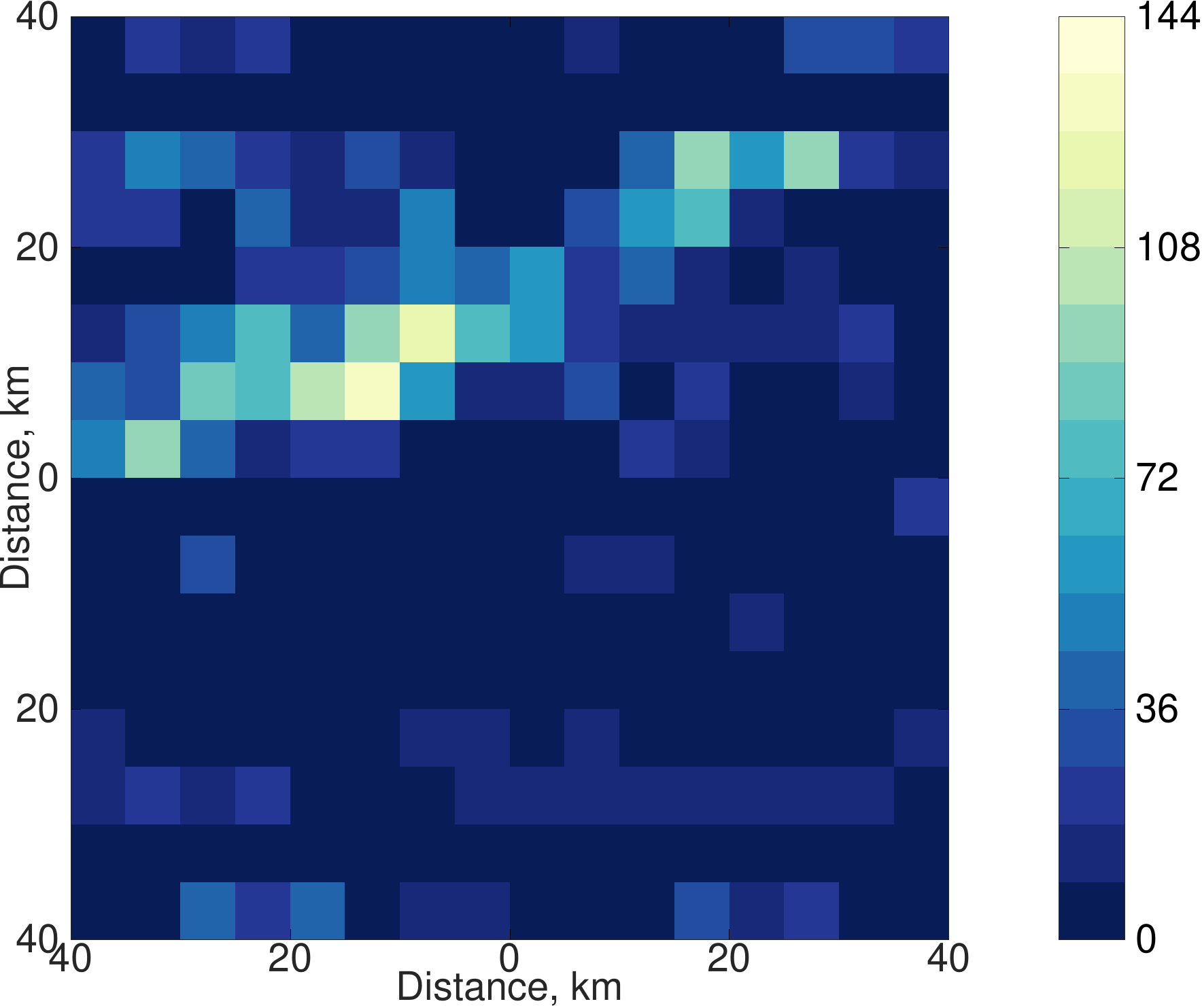}
	\end{center}
	\caption{Event recorded over the Pacific Ocean on September~7,
		2016.
		Left: waveforms of the brightest pixels of the PMT modules shown as
		rows 3--7 from the top of the snapshot.
		Bottom: a snapshot of the focal plane for time frame number~176.
		The bright arc started in the top NW corner of the
		focal plane and moved towards the SE corner with its
		brightness decreasing.
		}
	\label{fig:elve}
\end{figure}

A few other kinds of events registered with TUS are related to thunderstorm
activity and TLEs, among them so-called elves~\cite{1996GeoRL..23.2157F}.
Elves are short-lived optical events that manifest at the
lower edge of the ionosphere as bright rings expanding at the speed of
light up to a maximum radius of $\sim300$~km.  Up to December~16, 2016,
five events registered with TUS in the EAS mode have been identified as
elves. The first of them is shown in Fig.~\ref{fig:elve}.  The event was
registered on September~7, 2016, at 09:51:35 UTC over the Pacific Ocean
($11^\circ\!\!.62$S, $161^\circ\!\!.68$W).  It appeared as a bright arc
crossing the focal plane from North-West to South-East, fading steadily.
A lightning strike within 0.6~s was registered by the WWLLN at a
distance of about 180~km North-West from the centre of the TUS field of
view.  The position of the lightning, the form of the bright arc in the
focal plane, and the temporal evolution of the waveforms strongly
support the conjecture that this was an elve.  Another example of
an elve registered with TUS can be found in~\cite{TUS-ECRS-2016}.

\subsection{Events with noise-like waveforms}

In the majority of events registered by far with TUS ($>80\%$),
waveforms look like a stationary stochastic process with ADC counts
fluctuating around some average values that mostly depend on conditions
of observations.  Illumination of the focal surface is typically uniform
up to sensitivity of individual PMTs during moonless nights and far from
urban regions. Trigger of such events seems to be caused by random
fluctuations of the nocturnal UV background radiation.

\begin{figure}[!ht]
	\begin{center}
		\includegraphics[width=.44\textwidth]{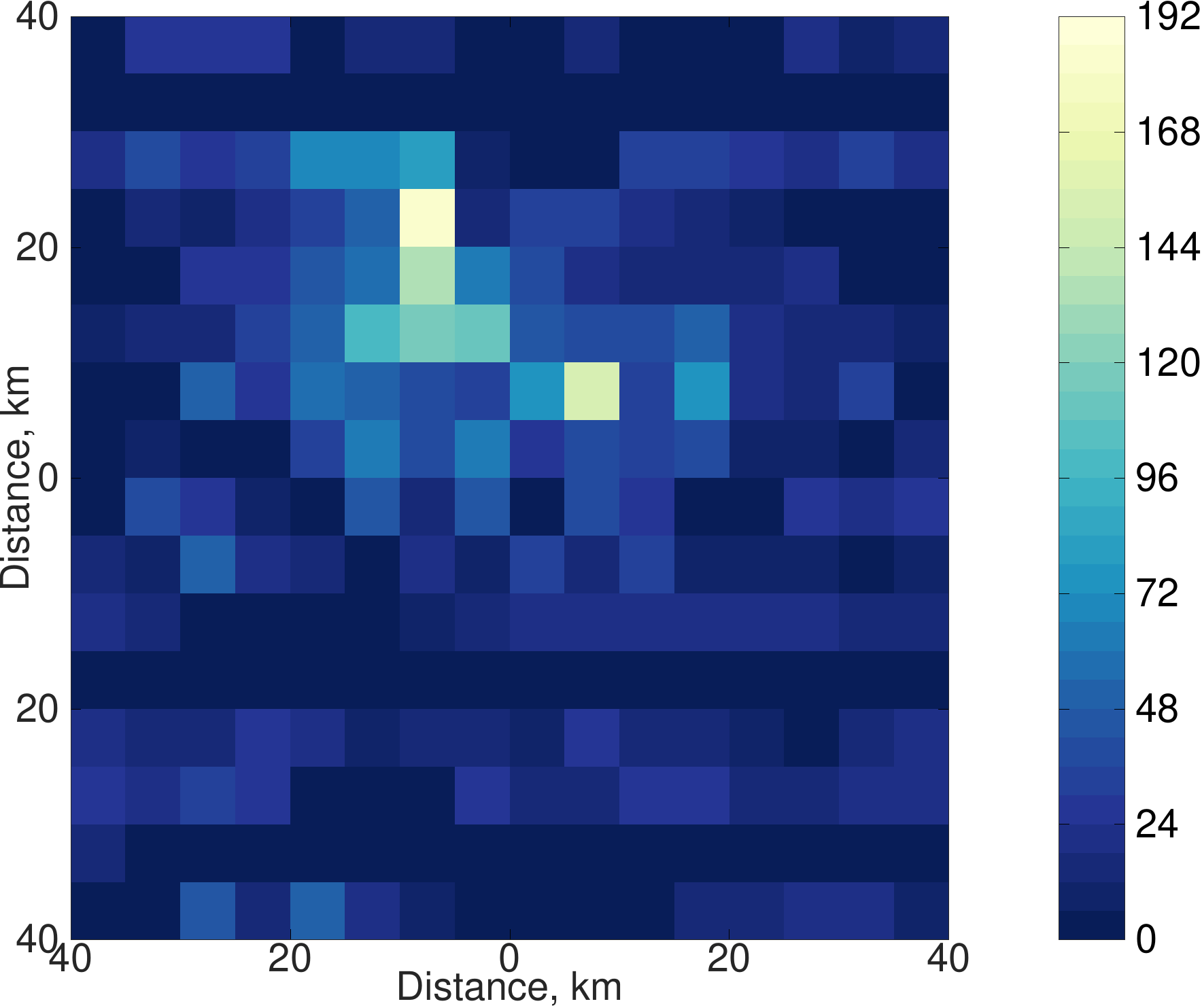}\qquad
		\includegraphics[width=.44\textwidth]{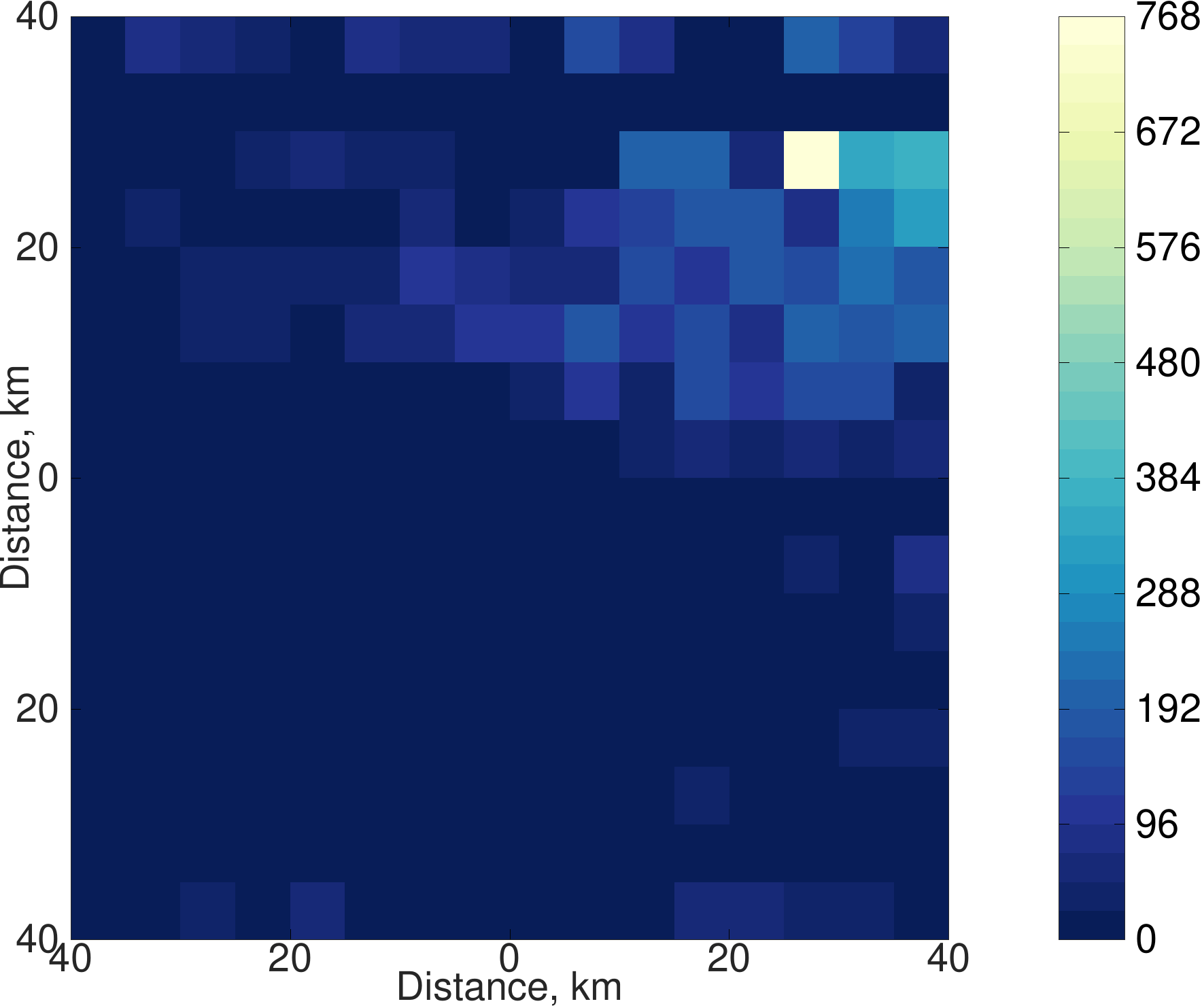}
	\end{center}
	\caption{Examples of snapshots of the focal plane
		made above cities. Left: Nagoya, Japan, in cloudy weather
		conditions.
		Right: Buenos~Aires, Argentina, in clear weather.
		}
	\label{fig:cities}
\end{figure}

Illumination of the focal plane becomes strongly non-uniform during
moonlit nights because TUS does not have side shields, and moonlight can
arrive directly at the focal surface. Other sources of non-uniform
illumination are auroral ovals, thunderstorm regions and industrial
sites, urban regions and generally objects related to human activities.
Fig.~\ref{fig:cities} presents examples of events recorded 
near Nagoya (September~28, 2016, at 14:22:47 UTC, centre of the FOV at
$35^\circ\!\!.1$N, $136^\circ\!\!.9$E) and Buenos~Aires (November~18,
2016, 02:43:49 UTC, $34^\circ\!\!.7$S, $58^\circ\!\!.9$W).
TUS routinely registers numerous events of this kind, especially in
regions with clear atmosphere, but in most cases
a concrete source of the bright signal cannot be identified because of
the insufficient spatial resolution of the instrument.
A clear frequency modulation at 100~Hz or 120~Hz is observed above
cities when TUS operates in the TLE mode with the time window of 0.4~ms,
see an example in~\cite{TUS-TEPA-2016}.  Interestingly, it was found
that some existing artificial sources (not related to the experiment)
can mimic strength and spatio-temporal dynamics of signals expected from
extensive air showers generated by EECRs.

There are also TUS events that do not fit in the above three groups.
Some of them form their own small subgroups, others are unique thus far.
An example of a violent flash registered by TUS on September~5, 2016,
near Sardinia, for which we have failed to identify an anthropogenic or
a natural source can be found in the slides of the
talk~\cite{UHECR2016-talk}.  At the time of writing, unusual events are
undergoing analysis and will be reported elsewhere.

\section{Conclusions}

TUS is the first orbital detector capable of registering fast UV flashes
in the nocturnal atmosphere of the Earth in the imaging mode, providing
much more detailed data than the earlier Tatiana, Tatiana-2 and Vernov
missions of Lomonosov Moscow State
University~\cite{2011CosRe..49..391G,Garipov2013,2016CosRe..54..261P}.
The results already obtained are mostly unexpected and provide important
information about the nocturnal UV background necessary for successful
development of the future KLYPVE (\mbox{K-EUSO}) and \mbox{JEM-EUSO}
missions.  TUS has also demonstrated that an orbital detector can be
utilized for simultaneously studying processes that manifest themselves
with UV flashes in the atmosphere but have inherently different nature
and time scales.  Still, following the main objective of the experiment,
much efforts are being put in the search for EAS candidates in the data.
A number of methods employed resulted in selecting a few promising
events.  Their detailed analysis is in progress, and its results will be
presented in a dedicated report.

\medskip

I would like to thank the organizing committee of UHECR2016 for the
invitation to take part in the inspiring conference, and personally
Shoichi~Ogio and Atsuko~Kitsugi for the help in organizing the visit.  I
thank my colleagues Boris~Khrenov and Pavel~Klimov for numerous
enlightening discussions.  My special thanks are addressed to a referee
for multiple suggestions improving the text.  The Lomonosov--UHECR/TLE
collaboration thanks Robert~Holzworth, the head of the World Wide
Lightning Location Network, for generously providing the data on
lightning strikes employed in the present study.  The work of the
Lomonosov--UHECR/TLE collaboration is supported from the Russian
Foundation for Basic Research grants No. 15-02-05498-a and No.
16-29-13065 in Russia and the National Research Foundation grants No.
2015R1A2A1A01006870 and No. 2015R1A2A1A15055344 in Korea.


\newpage
\noindent
\textbf{The Lomonosov--UHECR/TLE collaboration}

\bigskip
{\sloppy
\noindent
S.V.~Biktemerova$^{b}$,
A.A.~Botvinko$^{c}$,
N.P.~Chirskaya$^{a}$,
V.E.~Eremeev$^{a}$,
G.K.~Garipov$^{a}$,
V.M.~Grebenyuk$^{b,d}$,
A.A.~Grinyuk$^{a}$,
S.~Jeong$^{f}$,
N.N.~Kalmykov$^{a}$,
M.A.~Kaznacheeva$^{a}$,
B.A.~Khrenov$^{a}$,
M.~Kim$^{f}$,
P.A.~Klimov$^{a}$,
M.V.~Lavrova$^{a}$,
J.~Lee$^{f}$,
O.~Martinez$^{g}$,
M.I.~Panasyuk$^{a}$,
I.H.~Park$^{f}$,
V.L.~Petrov$^{a}$,
E.~Ponce$^{g}$,
A.E.~Puchkov$^{c}$,
H.~Salazar$^{g}$,
O.A.~Saprykin$^{c}$,
A.N.~Senkovsky$^{c}$,
S.A.~Sharakin$^{a}$,
A.V.~Shirokov$^{a}$,
A.V.~Tkachenko$^{b}$,
L.G.~Tkachev$^{b,d}$,
I.V.~Yashin$^{a}$,
M.Yu.~Zotov$^a$

}

\bigskip
\par\noindent{$^a$M.V. Lomonosov Moscow State University, GSP-1, Leninskie
	Gory, Moscow, 119991, Russia}
\par\noindent{$^b$Joint Institute for Nuclear Research, Joliot-Curie, 6,
	Dubna, Moscow region, Russia, 141980}
\par\noindent{$^c$Space Regatta Consortium, ul. Lenina, 4a,
	141070 Korolev, Moscow region, Russia}
\par\noindent{$^d$Dubna State University, University str., 19, Bld.1,
	Dubna, Moscow region, Russia}
\par\noindent{$^e$Department of Physics and ISTS, Sungkyunkwan
	University,	Seobu-ro 2066, Suwon, 440-746 Korea}
\par\noindent{$^f$Benem\'{e}rita Universidad Aut\'{o}noma de Puebla,
	4 sur 104 Centro Hist\'orico C.P. 72000, Puebla, Mexico}


\begin{thebibliography}{30}

\bibitem{Linsley-1e20-1963}
J.~Linsley, Phys. Rev. Lett. \textbf{10}, 146 (1963).

\bibitem{Dawson_etal-2017}
B.~R. {Dawson}, M.~{Fukushima}, and P.~{Sokolsky}, arXiv:1703.07897.

\bibitem{2015ApJ...804...15A}
A.~{Aab}, P.~{Abreu}, M.~{Aglietta}, et~al., Astrophys. J. \textbf{804}, 15
  (2015).

\bibitem{Tinyakov-ICRC-2015}
P.~{Tinyakov}, M.~Fukushima, D.~{Ikeda}, et~al., Proc. of Science (ICRC2015), ID326
  (2015).

\bibitem{1980BAAS...12Q.818B}
R.~{Benson} and J.~{Linsley}, {Bull. of the American Astron. Society}
		\textbf{12}, 818 (1980).

\bibitem{Benson-Linsley-1981}
R.~{Benson} and J.~{Linsley}, Proc. 17th Int. Cosmic Ray Conf., Paris,
  1981, Vol.~8, p.~145.

\bibitem{2001ICRC....2..831A}
V.~V. {Alexandrov}, D.~I. {Bugrov}, G.~K. {Garipov}, et~al., Proc. 27th
	Int. Cosmic Ray Conf., Hamburg, 2001, Vol.~2, p.~831.

\bibitem{TUS-ecrs2012}
B.~A. {Khrenov}, M.~I. {Panasyuk}, G.~K. {Garipov}, et~al., European
		Phys. J. Web of Conf. \textbf{53}, 09006 (2013).

\bibitem{TUS-expastron}
J.H.~Adams Jr., S.~Ahmad, J.-N. Albert, et~al., Exp. Astron. \textbf{40}, 315
  (2015).

\bibitem{2014NIMPA.763..604G}
A.~{Grinyuk}, M.~{Slunecka}, A.~{Tkachenko}, L.~{Tkachev}, P.~{Klimov}, and
  S.~{Sharakin}, Nucl. Instrum. Methods Phys. Res., Sect. A \textbf{763}, 604
  (2014).

\bibitem{1991JPhG...17..733L}
M.~A. {Lawrence}, R.~J.~O. {Reid}, and A.~A. {Watson}, J. of Physics G Nucl.
  Phys. \textbf{17}, 733 (1991).

\bibitem{AGASA-spectrum}
M.~{Takeda}, N.~{Hayashida}, K.~{Honda}, et~al., Phys. Rev. Lett. \textbf{81},
  1163 (1998).

\bibitem{Greisen-1966}
K.~Greisen, Phys. Rev. Lett. \textbf{16}, 748 (1966).

\bibitem{ZK-1966}
G.~T. {Zatsepin} and V.~A. {Kuz'min}, Sov. Phys. JETP Lett.,
  \textbf{4}, 78 (1966).

\bibitem{HiRes-GZK-2007}
D.~R. {Bergman} and {High Resolution Fly's Eye Collaboration}, Nucl. Phys. B
  Proc. Supp. \textbf{165}, 19 (2007).

\bibitem{Auger-GZK}
J.~{Abraham}, P.~{Abreu}, M.~{Aglietta},
  et~al., Phys. Rev. Lett. \textbf{101}, 061101 (2008).

\bibitem{klypve-2015}
G.~K. {Garipov}, M.~Yu. {Zotov}, P.~A. {Klimov}, M.~I. {Panasyuk}, O.~A.
  {Saprykin}, L.~G. {Tkachev}, S.~A. {Sharakin}, B.~A. {Khrenov}, and I.~V.
  {Yashin}, Bull. Rus. Acad. Sci. Physics \textbf{79}, 326 (2015).

\bibitem{JEM-EUSO-tool}
J.H.~Adams Jr., S.~Ahmad, J.-N. Albert, et~al., Exp. Astron. \textbf{40}, 19
  (2015).

\bibitem{TUS-sim-ApP-2017}
A.~{Grinyuk}, V.~{Grebenyuk}, B.~{Khrenov}, P.~{Klimov}, M.~{Lavrova},
  M.~{Panasyuk}, S.~{Sharakin}, A.~{Shirokov}, A.~{Tkachenko}, L.~{Tkachev},
  and I.~{Yashin}, Astropart. Phys. \textbf{90}, 93 (2017).

\bibitem{TUS-TEPA-2016}
P.~{Klimov}, B.~{Khrenov}, S.~{Sharakin}, M.~{Zotov}, N.~{Chirskaya},
  V.~{Eremeev}, G.~{Garipov}, M.~{Kaznacheeva}, M.~{Panasyuk}, V.~{Petrov},
  A.~{Shirokov}, and I.~{Yashin}, in
  \emph{Proc. of
  {I}nt. {S}ymposium {TEPA}-2016, {Y}erevan, {A}rmenia},
  ed. A.~{Chilingarian},
  (Yerevan Physics Institute, 2017), p.~122.

\bibitem{TUS-ECRS-2016}
S.~V. {Biktemerova}, A.~V. {Bogomolov}, V.~V. {Bogomolov}, et~al., 
  arXiv:1703.03738.

\bibitem{tracks-izvran}
P.~A. {Klimov}, M.~Yu. {Zotov}, N.~P. {Chirskaya}, B.~A. {Khrenov}, G.~K.
  {Garipov}, M.~I. {Panasyuk}, S.~A. {Sharakin}, A.~V. {Shirokov}, I.~V.
  {Yashin}, A.~A. {Grinyuk}, A.~V. {Tkachenko}, and L.~G. {Tkachev}, Bull. Rus.
  Acad. Sci. Physics \textbf{81}, 407 (2017).

\bibitem{Geant4}
S.~{Agostinelli}, J.~{Allison}, K.~{Amako}, et~al., Nucl. Instrum. Methods in
  Phys. Res., Sect. A \textbf{506}, 250 (2003).

\bibitem{2011BRASP..75..381S}
O.~P. {Shustova}, N.~N. {Kalmykov}, and B.~A. {Khrenov}, Bull. Russ. Acad.
  Sci., Physics \textbf{75}, 381 (2011).

\bibitem{2014AdSpR..53.1515B}
M.~{Bertaina}, S.~{Biktemerova}, K.~{Bittermann}, et~al., Adv. in Space Res.,
  \textbf{53}, 1515 (2014).

\bibitem{1996GeoRL..23.2157F}
H.~{Fukunishi}, Y.~{Takahashi}, M.~{Kubota}, K.~{Sakanoi}, U.~S. {Inan}, and
  W.~A. {Lyons}, Geophys. Research Lett. \textbf{23}, 2157 (1996).

\bibitem{UHECR2016-talk}
M.~{Zotov}, {URL}: https://indico.cern.ch/event/504078/contributions/2305948/;
  presentation at {UHECR}2016, {K}yoto, {J}apan (2016).

\bibitem{2011CosRe..49..391G}
G.~K. {Garipov}, P.~A. {Klimov}, V.~S. {Morozenko}, M.~I. {Panasyuk}, and B.~A.
  {Khrenov}, Cosmic Research \textbf{49}, 391 (2011).

\bibitem{Garipov2013}
G.~K. Garipov, B.~A. Khrenov, P.~A. Klimov, et~al., J. Geophys. Res.:
  Atmospheres \textbf{118}, 370 (2013).

\bibitem{2016CosRe..54..261P}
M.~I. {Panasyuk}, S.~I. {Svertilov}, V.~V. {Bogomolov}, et~al., Cosmic
  Research \textbf{54}, 261 (2016).

\end{thebibliography}
\end{document}